\documentclass[pra,twocolumn,superscriptaddress,floatfix]{revtex4}
\usepackage{amsmath}
\usepackage{amsthm}
\usepackage{bbm}
\usepackage{amsfonts}
\usepackage[colorlinks=true,citecolor=blue,linkcolor=blue,urlcolor=blue]{hyperref}
\usepackage[usenames]{color}
\usepackage{subeqnarray}
\usepackage{graphics}
\usepackage{subfigure}
\usepackage{dcolumn}
\usepackage{bm}
\usepackage{color}
\usepackage{verbatim}
\usepackage{epstopdf}
\usepackage{latexsym}
\usepackage{hyperref}
\usepackage{psfrag}
\usepackage{xcolor}
\usepackage{times}

\def\CC{\mathbbm{C}}
\def\Pr{\mathbb{P}}
\def\Hsym{\mathcal{H}_{\mathrm{sym}}^N}
\def\l{\lambda}
\def\vare{\varepsilon}

\begin{document}
\title{Entanglement Classification with Algebraic Geometry}
\author{M. Sanz}
\affiliation{Department of Physical Chemistry, University of the Basque Country UPV/EHU, Apartado 644, E-48080 Bilbao, Spain}
\author{D. Braak}
\affiliation{EP VI and Center for Electronic Correlations and Magnetism, University of Augsburg, 86135 Augsburg, Germany}
\author{E. Solano}
\affiliation{Department of Physical Chemistry, University of the Basque Country UPV/EHU, Apartado 644, E-48080 Bilbao, Spain}
\affiliation{IKERBASQUE, Basque Foundation for Science, Maria Diaz de Haro 3, 48013 Bilbao, Spain}
\author{I. L. Egusquiza}
\affiliation{Department of Theoretical Physics and History of Science, University of the Basque Country UPV/EHU, Apartado 644, 48080 Bilbao, Spain}

\begin{abstract} 
We approach multipartite entanglement classification in the symmetric subspace in terms of algebraic geometry, its natural language. We show that the class of symmetric separable states has the structure of a Veronese variety and that its $k$-secant  varieties are SLOCC invariants. Thus  SLOCC classes  gather naturally into families.  This classification presents useful properties such as a linear growth of the number of families with the number of particles, and nesting, i.e. upward consistency of the classification. We attach  physical meaning to this classification through the required interaction length of  parent Hamiltonians. We show that the states $W_N$ and GHZ$_N$ are in the same secant family and that, effectively, the former can be obtained in a limit from the latter.  This limit is understood in terms of tangents, leading to a refinement of the previous families. We compute explicitly the  classification  of symmetric states with \( N\leq4 \) qubits  in terms   of both secant families and  its  refinement using tangents.
This paves the way to further use of projective varieties in algebraic geometry to solve open problems in entanglement theory.
\end{abstract}

\date{\today}

\maketitle
\section{Introduction}
Entanglement is  a  cornerstone of quantum information science \cite{HHHH09}, and an essential resource for central quantum effects. Relevant instances of such effects are the teleportation of quantum states \cite{BBCJPW93,BPMEWZ97,RHRHRBLKBSKJB04,BCSBIJKLLOW04}, efficient information transmission throughout dense coding \cite{BW92}, as well as  codification of provably secure information via quantum cryptography \cite{BB84,GRTZ02}. Moreover, entanglement is widely considered a crucial  component for  quantum speed-up in both quantum computations and quantum simulations \cite{NC00,Sh97}, and it is now considered to play a role even in biology~\cite{MRLA-G08} and biomimetics~\cite{A-RSLS14,A-RSLS16}. Therefore, the characterisation, quantification, and classification of entanglement are crucial milestones in quantum information.

In spite of its relevance, there is no general quantification of entanglement for many-body systems. Rather,  quantum states connected by stochastic local operations assisted with classical communication (SLOCC), which perform probabilistically the same quantum tasks, can be collected into entanglement classes, also known as SLOCC classes~\cite{DVC00}. Even so, there is an infinite number of these classes for four or more parties that may be gathered, in turn, into a finite number of entanglement families~\cite{VDdMV02, LLSS06, LLSS07, BKMGLS09, WDGC13}. The challenge consists in providing classifications into families with mathematical and physical relevance. In this article, we consider the problem of entanglement classification in the frame of algebraic geometry~\cite{Brody200119,PhysRevA.67.012108,jmp10.1063/1.4753989}. Extending previous works~\cite{BGH07}, we show that the class of symmetric separable states corresponds to a Veronese variety, and that the $k$-secant varieties constructed from it are SLOCC invariants. Moreover, the $k$-secant variety to which a state belongs determines the maximum rank of the reduced density  matrices of this state and therefore, by a construction similar to \cite{MEdCSLS15},  the interaction length of a Hamiltonian having this state as a ground state, thus establishing a deep connection with condensed matter. As an example of this connection, we prove that, if a given state is a ground state of a Hamiltonian with interaction length $k$, then with certainty it will belong to the $k$-secant variety. Similarly, we know that every state in the $k$-secant variety can be constructed or simulated by a Hamiltonian with interaction length  $k$. In fact, if the state belongs to the \( k \)-secant variety but not to the \( k-1 \)-secant variety, then it will be a ground state only for Hamiltonians with interaction length equal to or larger than \( k \). Altogether this provides a deep physical insight to our classification. Finally, we study in detail the states $W_N$ and GHZ$_N$, showing that they are in the same family and that, effectively, the former can be obtained in a limit from the latter, providing a constructive procedure to approach them with vanishing probability.

In the following, we confine our analysis  to the \emph{symmetric} subspace of a system of  $N$  identical parties, each of them described with a finite-dimensional Hilbert space $\mathcal{H}\simeq \mathbbm{C}^d$. We shall denote the symmetric subspace of \( \mathcal{H}^{\otimes N} \) by  $\mathrm{Sym}\left[ \mathcal{H}^{\otimes N} \right] =: \Hsym$. Each element of $\Hsym$ is invariant under the action of $S_N$, permuting the factors in each of its tensor components.
Normalizing the state vectors corresponds as usual to the
projective representation in $\mathbb{P}\left(\Hsym\right)$. The dimension of $\Hsym$ reads
\begin{equation}
\mathrm{dim}\left[\Hsym\right]={{N+d-1}\choose{N}}\,.
\end{equation}
For qubits ($d=2$), this is $N+1$. The manifold  $\mathbb{P}\left(\Hsym\right)$
has complex dimension $\mathrm{dim}\left[\Hsym\right]-1$. 
Once a basis \( \{\bm{e}_j\}_{j=1}^d \) has been selected,
an orthogonal  basis in $\Hsym$ (called the induced basis) is given by the vectors
\begin{equation}
|[n_0,\dots,n_{d-1}]\rangle =
\sum_{J\in F[n_0,\ldots,n_{d-1}]} \bm{e}_{J(1)}\otimes\ldots\otimes \bm{e}_{J(N)},
\end{equation}
where the functions $J$:$\{1,\ldots,N\}\rightarrow\{0,\ldots,d-1\}$ are in
the set $F[n_0,\ldots,n_{d-1}]$ if they take $n_j$ times the value $j$. The indices $0\le n_j\le N$ fulfill $\sum_{k=0}^{d-1}n_k=N$. The vectors 
$|[n_0,\dots,n_{d-1}]\rangle$ are not normalized but are proportional to the normalized Dicke states for qudits:
\begin{equation}
|D^{(N)}_{[n_0,\dots,n_{d-1}]}\rangle =
\left(\frac{N!}{n_0!\ldots n_{d-1}!}\right)^{-1/2}|[n_0,\dots,n_{d-1}]\rangle.
\end{equation}
For two qubits ($N=d=2$), the induced basis reads
$|[2,0]\rangle=|00\rangle$, $|[1,1]\rangle=|01\rangle+|10\rangle$, 
$|[0,2]\rangle=|11\rangle$.

\section{Entanglement Classification via Algebraic Varieties}

We shall show in the following that the entanglement classification via SLOCC invariance in $\Hsym$ is related to well-known concepts of algebraic geometry. We begin with the simplest SLOCC class, the separable states, to afterwards construct higher entanglement families and show that they are SLOCC invariant.
\subsection{Separable States and Veronese Variety}
Each separable state \(\psi_N\) in $\mathbb{P}(\Hsym)$ issues from a state  $\psi\in\mathbb{P}(\CC^d)$. If the vector representing the state \( \psi \) reads $|\psi\rangle=\sum_{j=0}^{d-1}x_j\bm{e}_j$,  the vector representing \( \psi_N \) reads
\begin{equation}
|\psi_N\rangle=\sum_{[n_0,\ldots,n_{d-1}]}\prod_{j=0}^{d-1}x_j^{n_j}|[n_0,\ldots,n_{d-1}]\rangle,
\end{equation}
i.e. the coordinates of $\psi_N$ in the induced basis are homogeneous monomials of degree $N$ in the $d$ variables $x_j$. In  the projective representation, the homogeneous coordinates of $\psi\in \Pr(\CC^d)=\Pr^{d-1}$ are $[x_0:x_1:\ldots:x^{d-1}]$. Likewise, the homogeneous coordinates of $\psi_N\in \Pr(\Hsym)$ read $[x_0^N:x_0^{N-1}x_1:x_0^{N-1}x_2:\ldots:x_{d-1}^N]$.

We therefore have a map $\nu_N$ from $\Pr^{d-1}$ to $\Pr(\Hsym)$,
\begin{equation} 
\nu_N(\psi)=\psi_N
\label{veronese}
\end{equation}
which is nothing else but the classical \emph{Veronese map} in algebraic geometry \cite{Ha92}. The image of $\Pr^{d-1}$ under the Veronese map is an algebraic variety with complex dimension $d-1$, the Veronese variety $V^{N,d-1}=\nu_N(\Pr^{d-1})$. 
For qubits, it is a \emph{rational normal curve}. 
It can be characterized by the condition that the homogeneous coordinates 
$[z_{[N,0\ldots,0]}:z_{[N-1,1\ldots,0]}:\ldots :z_{[0,\ldots,N]}]$ of each of its points satisfy the quadratic relations $z_Iz_J=z_Kz_L$ if $n_j(I)+n_j(J)=n_j(K)+n_j(L)$ for each $0\le j\le d-1$.

The Veronese map is injective, therefore the correspondence between symmetric separable states $\psi_N$ and points $p\in V^{N,d-1}$ is one to one. The crucial point is now that the Veronese variety is a natural SLOCC invariant. Let $A$ be a SLOCC transformation of the state $\psi\in \mathbb{P}(\CC^d)$. \( A \) is an element of the projective group, and it is represented by $A\in GL(d,\CC)$ (in an abuse of notation that should not lead to confusion),  acting on
$\Hsym$ as $A^{\otimes N}$. It therefore intertwines with the Veronese map,
\begin{equation}
A^{\otimes N}(\nu_N(\psi))=\nu_N(A(\psi)),
\label{intertw}
\end{equation}  
which means that a SLOCC transformation amounts simply to a reparametrization of
$V^{N,d-1}$ in terms of $\Pr^{d-1}$.
This fact allows us to derive other SLOCC invariants, corresponding to entangled states, by constructing other algebraic varieties from  $V^{N,d-1}$ which  in this way inherit its SLOCC invariance.

\subsection{Entanglement Families as Secant Varieties}

A natural way to introduce entanglement is to write an entangled symmetric state $\phi\in \mathbb{P}(\Hsym)$ in terms of a linear combination of $k$ separable symmetric vectors,
\begin{equation}
|\phi\rangle=\sum_{j=1}^kc_j|\psi_N^{(j)}\rangle,
\end{equation}
where the $\psi_N^{(j)}$ (viewed as elements of $\Pr(\Hsym)$) are $k$ different points on $V^{N,d-1}$. Given a state \(\phi\), the smallest number \( k \) such that it can be represented this way is called its symmetric tensor rank in the mathematical literature, and Schmidt rank or Schmidt measure in the physical literature (albeit not usually restricted to the symmetric case). Given two states related by a SLOCC transformation, their symmetric tensor rank is the same. Thus symmetric tensor rank is a SLOCC invariant. Let us denote the set of states with symmetric tensor rank smaller than or equal to \( k \) as  \( \sigma_k^*(V^{N,d-1}) \). Clearly, the sets  \( \sigma_k^*(V^{N,d-1}) \) are SLOCC invariant. We say that a state \(\phi\) is a proper $k$-secant state if it belongs to \( \sigma_k^*(V^{N,d-1}) \) but not to  \( \sigma_{k-1}^*(V^{N,d-1}) \), i.e. if its symmetric tensor rank is \( k \). We say that it is a proper secant state if it belongs to \( \sigma_k^*(V^{N,d-1}) \) for some \( k \).

Crucially for our purposes, \( \sigma_k^*(V^{N,d-1}) \) is not closed. However, its closure with respect to the topology induced from $\Hsym$ \cite{Note0} is again an algebraic variety, the $k$-secant variety $\sigma_k(V^{N,d-1})$ of the Veronese variety \cite{Ha92}. We shall therefore define the secant entanglement families as follows:

A state $\phi\in \Pr(\Hsym)$ belongs to the $k$-th entanglement family if it is an element of the $k$-secant variety but not of the $(k-1)$-secant variety.

Because $V^{N,d-1}$ is not a subset of any hypersurface in $\Pr(\Hsym)$, every state in $\Pr(\Hsym)$ is located at least in $\sigma_{ \mathrm{dim}[\Hsym]}(V^{N,d-1})$, but the maximal $k$ needed to span the whole space is in fact much lower.  Simplifications arise for qubits, in which case the Veronese variety $V^{N,1}$ is a rational normal curve and $\mathrm{dim}[\Pr(\Hsym)]=N$. Then we have \cite{Note2}
 \begin{equation}\label{eq:nondefective}
\mathrm{dim}\left[ \sigma_k\left( V^{N,1}\right)\right]= \mathrm{min}\left(2k-1,N\right)\,.
\end{equation}
Therefore, the $k$-secant variety coincides with the total projective space $\Pr^N$ when $k = \lfloor \frac{N}{2}\rfloor+1$. 
A related partial result for qudits is given by the Alexander-Hirschowitz Theorem \cite{J.:1995hl}. 

\subsection{SLOCC invariance of the Entanglement Families}

The $k$-secant varieties $\sigma_k(V^{N,d-1})$ are SLOCC invariants. This follows from the injectivity of the Veronese map, the fact that the local operator $A$ is invertible and the intertwining property \eqref{intertw}. We consider first proper secants located in  $\sigma^*_k$. If the $k$ points $p_j\in V^{N,d-1}$ are independent (i.e. they define generic elements of $\sigma^*_k\backslash \sigma_{k-1}^*$), their images $A^{\otimes N}(p_j)\in V^{N,d-1}$ are independent as well and their linear span is therefore again in $\sigma^*_k\backslash \sigma_{k-1}^*$. 
This fact is easy to prove for qubits, because any $N+1$ points on the rational normal curve 
$\nu_N(\Pr^1)$ are independent \cite{Ha92} (see also a proof in \cite{MEdCSLS15}). The images of the $k$ points $p_j\in V^{N,1}$ are therefore independent if they are {\emph{different}, which is guaranteed by the injectivity of $\nu_N$ and $A$. 
We shall define the elements of the tangent manifolds forming the closure
of $\sigma_k^*$ by a limit procedure in the induced standard topology (see below). 
Because $A$ is then (trivially) continuous, the result for proper secants carries over to the tangent varieties of $\sigma^*_k$, which are mapped onto themselves under a SLOCC transformation. We conclude  that the property of an arbitrary entangled symmetric state of $N$ qudits to be the linear combination of exactly $k$ separable \emph{symmetric} states is a SLOCC invariant feature. As such it concerns only  elements of $\sigma^*_k$, for example the $\mathrm{GHZ}_N$-states. There are, of course, symmetric entangled states which are linear combinations of \emph{non-symmetric} separable states, for example the $W_N$-states. These states are elements of $\Pr(\Hsym)$ and must therefore belong to some secant variety $\sigma_k$ of $V^{N,1}$. However, they are not proper elements but points in one of the 
tangent varieties of $V^{N,1}$, as \emph{limits} of proper elements in 
$\sigma^*_k(V^{N,1})$. This provides us with a more detailed geometric description of entanglement beyond the grouping into families. The fact that tangents are limits of secants entails \emph{asymptotic SLOCC equivalence} between some states which belong to different SLOCC classes but to the same family. A detailed discussion of  asymptotic equivalence between the $\mathrm{GHZ}_N$ and $W_N$-states will be given below.

We confine ourselves in the following to qubits with $\Pr(\Hsym)=\Pr^N$. In this case, the secant varieties $\sigma_k(V^{N,1})$ can be characterized as 
\emph{determinantal varieties}. Let an element $\phi$ of $\Pr^N$ be given by its
homogeneous coordinates $[c_0:c_1:\ldots :c_N]$ (the $c_j$ are the coordinates of the corresponding vector in $\Hsym$, written in the induced basis) and consider for $1\le j\le N-1$ the \emph{catalecticant matrix} 
\cite{Note3,Sy52}:
\begin{equation}\label{eq:catalecticant}
C_j(\phi)=
 \begin{pmatrix}
c_0& c_1&\cdots& c_j\\
c_1& c_2&\cdots&c_{j+1}\\
\vdots&\vdots&\ddots&\vdots\\
c_{N-j}&c_{N+1-j}&\cdots&c_N
\end{pmatrix}\,.
\end{equation}

In general, a rank $k$ determinantal variety $X_k(M)\subset \Pr^N$ is defined by the condition that a $m\times n$ matrix $M(\phi)$, with $\phi\in \Pr^N$ has rank $k$ or less for all $\phi\in X_k(M)$.   
A standard theorem of algebraic geometry \cite{Note4} states now that the
rank $k$ determinantal variety defined via $C_j$ in \eqref{eq:catalecticant} 
is just the $k$-secant variety $\sigma_k(V^{N,1})$, if $k\le j$. Because the maximal possible  rank of $C_j(\phi)$ is $\lfloor \frac{N}{2}\rfloor +1$, we recover the result from \eqref{eq:nondefective} for the maximal $\sigma_k(V^{N,1})$. The separable states (located on the Veronese curve itself) correspond to $C_j(\phi)$ with rank one. This is easy to see, because each point $\phi$ on $V^{N,1}$ has homogeneous coordinates $[x_0^N : x_0^{N-1}x_1 : \ldots : x_1^N]$ , or $c_j=x_0^{N-j}x_1^j$. When plugged into \eqref{eq:catalecticant}, each row is a multiple of the first, therefore $C_j(\phi)$ has rank one.  

\subsection{Asymptotic Equivalence Between $\text{GHZ}_{N}$ and $W_N$ States}\label{b-rank}

We shall now look at the GHZ$_N$ and $W_N$ states, 
and compute the maximal rank of their catalecticant matrices. Let us recall the definition of the corresponding vectors
\begin{subequations}
\begin{align*}
|\text{GHZ}_N\rangle & = \frac{1}{\sqrt{2}} \left( |0\rangle^{\otimes N} + |1\rangle^{\otimes N}\right) \\
& = \frac{1}{\sqrt{2}} \left( |D^{(N)}_{[N,0]}\rangle + |D^{(N)}_{[0,N]}\rangle\right), \\
|W_N\rangle & = \frac{1}{\sqrt{N}}(|10\ldots 0\rangle + \cdots +|00\ldots1\rangle) = |D^{(N)}_{[N-1,1]}\rangle, 
\end{align*}
\end{subequations}
In order to give examples for the classification of the four-qubit case, let us also add, for \( N\geq4 \), the definition of a third well-known inequivalent for $N\geq4$ class of states, namely $|X_N\rangle$~\cite{MEdCSLS15,siewert10}, 
\begin{equation}
|X_N(z)\rangle = (N-1)\left(|1\rangle^{\otimes N} + z^{N-1}\sqrt{N}|W_N\rangle\right),
\end{equation}
up to normalization. 

Now we can state that for all \( N\geq2 \) both the GHZ$_N$  and  \( W_N \) states belong to \( \sigma_2(V^{N,1}) \). On the other hand, for \( N\geq4 \), the \( X_N \) states belong to  \( \sigma_3(V^{N,1}) \). Indeed, GHZ$_N$ states have non-zero $c_0 = c_N = 1$, so the rank of all $C_j$ is 2. $W_N$ states have only one non-zero coefficient, namely $c_1 = 1$, and all the ranks are trivially $2$. Both types of states belong therefore to  $\sigma_2(V^{N,1})$. As to \( X_N \) states, they present non-zero \( c_1 \) and \( c_N \), with all other coefficients zero. Thus \(  \mathrm{rank}\left[C_2(X_N)\right]=3 \), with no higher rank for other \( C_j(X_N) \). It follows that \( X_N\in \sigma_3(V^{N,1}) \).

Obviously, the classification of symmetric states using $k$-secant varieties  is thus equivalent with the characterization of elements $\phi \in\Pr(\Hsym)$ via their 
\emph{symmetric tensor border rank} (stbr), which is defined as the maximal rank of $C_j(\phi)$  for $j\le N/2$ \cite{LT10}. Like the symmetric tensor rank, Schmidt measure \cite{PhysRevA.64.022306} and bond dimension \cite{MEdCSLS15}, it is a tensor rank, a measure of the minimal number of components in a tensor decomposition. We conclude that the stbr gives the number of separable states appearing in a 
\emph{symmetric} tensor decomposition of $\phi$, if it exists, or, alternatively, the number of separable states in a symmetric tensor decomposition that, in a certain limit, produce \(\phi\). 
 
 \section{Two- and Three-Qubit States Classification}
 
We study now the case of two and three qubits in detail. In both cases, the second secant variety $\sigma_2(V^{N,1})$ spans $\Pr^N$. For $N=2$, the Veronese curve consists of the points $[1:z:z^2]\cup [0:0:1]$. Then 
\begin{equation}
\sigma^*_2=[\lambda+\mu:\lambda z+\mu w:\lambda z^2 +\mu w^2]
\cup [1:z:z^2+\mu]\, .
\end{equation}
It is easy to see that both $|\mathrm{GHZ}_2\rangle=[1:0:1]$ and $|W_2\rangle=[0:1:0]$ are proper elements of $\sigma^*_2(V^{2,1})$. 

A proper element $p^s(p_0,p_1;\l)$ of $\sigma^*_2(V^{N,1})$ not in the Veronese variety \( V^{N,1} \)  can be written as a point on the secant line $s(p_0,p_1)\subset \sigma_2^*(V^{N,1})$, represented by  $|p_0\rangle+\lambda(|p_1\rangle-|p_0\rangle)$, where $|p_0\rangle$ and $|p_1\rangle$ are vector representatives of {\em different} points on $V^{N,1}$, \( p_0 \) and \( p_1 \) (later we will slightly abuse notation by denoting elements of the secant line as \( p=p_0+ \lambda(p_1-p_0) \), whereby we mean the construction presented here). Notice that 
$\sigma_2^*(V^{N,1})$ is not closed in general, i.e. for generic \( N \).

Let us now consider the limit $p_1\rightarrow p_0$, which gives the tangent to $V^{N,1}$ at $p_0$. Because all elements of $V^{N,1}$ except $[0:0:\ldots :1]$ can be written as
$p_j=[1:z_j:z_j^2:\ldots :z_j^N]$, we write $z_1(\varepsilon)=z_0+\varepsilon$. An element 
$p^t(p_0;\l)$ of the tangent space at $p_0$ reads therefore (with the slight abuse of notation mentioned above)
\begin{align}
p^t(p_0;\l) &=\lim_{\varepsilon\rightarrow 0} \left[p_0+\frac{\lambda}{\varepsilon}(p_1(\varepsilon)-p_0)\right]  \label{tangent}\\
=[1:z_0+&\lambda : z_0^2+2\lambda z_0 : z_0^3+3\lambda z_0^2 :\ldots:z_0^N+N\lambda z_0^{N-1}]  \nonumber\,.
\end{align} 
To obtain the point on the tangent whose first homogeneous coordinate is zero, we write
\begin{equation}
\tilde{p}^t(p_0)=\lim_{\varepsilon\rightarrow 0} \frac{1}{\varepsilon}(p_1(\varepsilon)-p_0) .  \label{tangent2}
\end{equation}
 For $N=2$ we have the special situation that all points on the tangents,
$p^t=[1:z_0+\lambda:z_0^2+2\lambda z_0]$, respectively 
$\tilde{p}^t=[0:1:2z_0]$, lie also on proper secants. Namely, 
\begin{align}
p^t(p_0;\l)=& [1:z_0+\lambda:(z_0+\lambda)^2]-\lambda^2[0:0:1] \nonumber\\
\tilde{p}^t(p_0)=& [1:u:u^2]-[1:w:w^2]\, ,
\end{align}
with $u=z_0+1/2$, $w=z_0-1/2$. That means that 
\emph{all} elements of $\Pr^2$ are proper elements of $\sigma_2^*(V^{2,1})$. The set \( \sigma_2^*(V^{2,1}) \) is closed, i.e. 
$\overline{\sigma_2^*(V^{2,1})}= \sigma_2^*(V^{2,1})=\sigma_2(V^{2,1})$.
All entangled symmetric states of two qubits are linear combinations of two symmetric separable states. The two entanglement families coincide with the two entanglement classes.

This is no longer the case for $N=3$. The tangent points defined in \eqref{tangent}, \eqref{tangent2} cannot be expressed as proper elements of $\sigma_2^*(V^{3,1})$, which spans all of $\Pr^3$ only if the tangent variety is included as its closure.
We consider now the tangent to $p_0=[1:0:0:0]$,  since all others can be mapped to it by a SLOCC transformation. It follows that
\begin{align}
p^t(p_0;\lambda)=& [1:\lambda:0:0]\, , \nonumber\\
\tilde{p}^t(p_0)=& [0:1:0:0]\, .
\end{align}
We call the union of these points (for all $p_0\in V^{3,1}$) the second tangent variety 
$\tau_2(V^{3,1}) =\overline{\sigma_2^*(V^{3,1})}\setminus \sigma_2^*(V^{3,1})$.

One recognizes immediately $\tilde{p}^t$ as the state $W_3$. The state 
$p^t(\lambda)$ can be obtained from $W_3$ with the SLOCC transformation
\(A= \left(\begin{array}{cc}1/3&0\\ 1&\lambda\end{array}\right)\) (the vector \( |0\rangle \) is represented as \( \left(\begin{array}
{cc}0&1
\end{array}\right)^T \)).
 
The state $\mathrm{GHZ}_3$ is a proper point of $\sigma^*_2(V^{3,1})$. It follows that there are three entanglement classes for three qubits, namely elements of $V^{3,1}$ (separable), $\sigma^*_2(V^{3,1})\backslash V^{3,1}$ (symmetric superposition, $\mathrm{GHZ}$) and 
$\tau_2(V^{3,1})$ (asymmetric superposition, $W$). As it is a point of the tangent to $[1:0:0:0]=\left[|0\rangle^{\otimes 3}\right]$, the state $W_3$ can be obtained \emph{asymptotically} from $\mathrm{GHZ}_3$, employing the singular SLOCC transformation
 \(A_\varepsilon= \varepsilon^{-1/3}\left(\begin{array}{cc}0&\varepsilon \\ -1&1\end{array}\right)\),
\begin{equation}
|W_3\rangle=\lim_{\varepsilon\rightarrow 0}A_\varepsilon^{\otimes 3}|\mathrm{GHZ}_3\rangle\, .
\end{equation}
W. D\"urr {\it et al} \cite{DVC00} perform parameter counting in their analysis of the three-qubit states and thus remark that the $W_3$ state class is not dense in the space of states, while the $\mathrm{GHZ}_3$ class has the adequate number of parameters. Here, we are providing an explicit construction of that fact (for other perspectives regarding this limit see for instance \cite{PhysRevLett.101.140502,jmp1.4908106,PhysRevLett.112.160401}; for a detailed analysis in algebraic geometry note Ref. \cite{1751-8121-49-8-085201}). \\

We now explicitly show the stochastic implementation of the asymptotic generation of \( W_3 \) from GHZ$_3$, by first reminding the reader of the stochastic implementation of a general transformation. Given an invertible matrix \( A \), define the operators \( E= A/ \sqrt{\|A^{\dag}A\|} \) and \( \bar{E}= \sqrt{1- E^{\dag}E} \), which will be associated with a measurement with two possible outcomes, \( \alpha \) and \( \bar{\alpha} \). Starting from a general single qubit state \( \psi \), the outcome \( \alpha \) will appear with probability \(p(\alpha|\psi)= \langle\psi|E^{\dag}E|\psi\rangle \), and following the generalised von Neumann-L\"uders rule, the system will be in a state represented by the normalised vector \( E|\psi\rangle / \sqrt{p( \alpha|\psi)} \) if indeed the outcome \(\alpha\) has been obtained.  Correspondingly,  \(p(\bar{\alpha}|\psi)= \langle\psi|\bar{E}^{\dag}\bar{E}|\psi\rangle \), and  \( \bar{E}|\psi\rangle/ \sqrt{p(\bar{\alpha}|\psi)} \). We say that the state \( A\psi \) has been obtained if the outcome of the measurement is \(\alpha\), and that the success probability for that state is \( p(\alpha|\psi) \).

For the case of interest, we see that the state \( A_\epsilon^{\otimes 3} ( \mathrm{GHZ}_3)\) will be obtained with success probability 
\begin{equation}
p=  \frac{4 \epsilon^2\left(3+3 \epsilon^2+ \epsilon^4\right)}{\left(2+ \epsilon^2+ \sqrt{4+ \epsilon^4}\right)^3} = \frac{3}{16} \epsilon^2 + O\left( \epsilon^4\right)
\end{equation}
so, of course, the probability tends to $0$ as $\varepsilon \rightarrow 0$, as was only to be expected since \( W_3 \) and \( \mathrm{GHZ}_3 \)  belong to different SLOCC classes. 

\section{Four-Qubit States Classification}

One sees from \eqref{tangent2} that the $W_N$-states are elements of the tangent lines obtained as limits from $\sigma^*_2(V^{N,1})$ for all $N$. Likewise, the states $\mathrm{GHZ}_N$ are proper elements of  $\sigma^*_2(V^{N,1})$. However, the higher $k$-secant varieties allow for many different limit procedures, leading to tangent $(k-1)$-planes of various kinds. It follows from our general argument above that each of them constitutes a \emph{different} SLOCC invariant.      

Let us consider thus $N=4$ and construct the corresponding classification. The second secant variety $\sigma_2(V^{4,1})$ has dimension $3$, and it is again classified according to generic secant points (as the $\mathrm{GHZ}_4$ state) and tangent points (including $W_4$). The third secant variety, $\sigma_3(V^{4,1})$, has dimension $4$ and is isomorphic to the whole projective representation $\Pr^4$ of the symmetric space ${\cal H}^4_{\textrm{sym}}$ according to \eqref{eq:nondefective}. 

For proper 3-secant elements, i.e.  in $\sigma_3^*(V^{4,1})\backslash \sigma_2^*(V^{4,1})$, we choose three different points $p_0\neq p_1\neq p_2\in V^{4,1}$. The set $\sigma_3^*\backslash \sigma_2^*$ is the union of the secant planes $s(p_0,p_1,p_2)$, consisting of points
\begin{equation}
p^s(p_0,p_1,p_2;\l,\mu)=p_0+\lambda(p_1-p_0)+\mu(p_2-p_0)\, ,
\label{secant3}
\end{equation}
with $\l,\mu\neq 0$.
As emphasized before, none of these points is located on a secant line $s(p_0',p_1')\in\sigma_2(V^{4,1})$, apart from the trivial case $\{p_0',p_1'\}\subset\{p_0,p_1,p_2\}$, because any five points of $V^{4,1}$ are independent. Therefore $\sigma_3^*$ describes four-qubit states which are a linear combination of three symmetric separable states. 

To construct the closure of $\sigma_3^*$, we consider the tangent planes of $\sigma_3^*$ at $p_0=[1:0:0:0:0]$, equivalent to the generic case. There are two types of them:

1.)\ \ The point $p_1$ approaches $p_0$, $p_1\rightarrow p_0$, while $p_2\neq p_0$ is kept fixed. We find
\begin{equation}
p^t(p_0,p_2;\l,\mu)=\lim_{\vare\rightarrow 0} \, p_0+\frac{\l}{\vare}(p_1(\vare)-p_0) +\mu(p_2-p_0)\, ,
\end{equation}
respectively
\begin{equation}
 \tilde{p}^t(p_0,p_2;\mu)=\lim_{\vare\rightarrow 0} \, \frac{1}{\vare}(p_1(\vare)-p_0) +\mu(p_2-p_0)\, .
\end{equation}  
(compare Eqs. \eqref{tangent}, \eqref{tangent2}). This means that the tangent planes can be written in terms of tangent lines as follows,
\begin{align}
p^t(p_0,p_2;\l,\mu)=& p^t(p_0;\l) + \mu(p_2-p_0) \, ,\nonumber\\
\tilde{p}^t(p_0,p_2;\mu)=& \tilde{p}^t(p_0)+\mu(p_2-p_0)\,.
\end{align}
We denote the union of these points as the tangent variety $\tau_3(V^{4,1})$.
For example, the state $X_4=[0:1:0:0:w]$ is an element of $\tau_3$. We have
$p^t(p_0;w^{-1})=[1:w^{-1}:0:0:0]$. With $p_2=[0:0:0:0:1]$ and $\mu=1$, one finds
\begin{align}
X_4 =& [1:w^{-1}:0:0:0]+[-1:0:0:0:1] \nonumber\\
\phantom{X_4}=& p^t(p_0,p_2;w^{-1},1)\, .
\end{align}
We conclude that $X_4$ is a member of a new entanglement class, as it is 
neither in $\sigma_3^*(V^{4,1})$ (it cannot be written in terms of three symmetric separable states) nor in $\sigma_2(V^{4,1})$, as  GHZ$_4$ and $W_4$ indeed are.  It can be obtained asymptotically from a generic element of $\sigma_3^*$, but {\em not} from GHZ$_4$, because $ \textrm{GHZ}_4 \in \sigma_2^*$.

2.)\ \ The second type of tangent plane is obtained by letting both $p_1$ and $p_2$ approach $p_0$.
The result is
\begin{align}
\bar{p}^t(p_0;\l,\mu) =&
\lim_{\vare, \vare'\rightarrow 0}\, 
p_0+\frac{\l}{\vare}(p_1(\vare)-p_0)+\frac{\mu}{\vare'}(p_2(\vare')-p_0)\nonumber\\
 =& [1:\l+\mu:0:0:0]=p^t(p_0;\l+\mu)\, ,
\end{align}
with $p_j=[1:z_j:z_j^2:z_j^3:z_j^4]$ and $z_{0,1,2}=0,\vare,\vare'$.
The resulting points are located on the tangents to $V^{4,1}$ and are therefore elements
of $\sigma_2$, which means that the tangent planes of the first type exhaust the points in the closure of $\sigma_3^*$. We find finally
\begin{equation}
\Pr^4=\sigma_3(V^{4,1})=\overline{\sigma_3^*(V^{4,1})}
=\sigma_3^*(V^{4,1})\cup\tau_3(V^{4,1})\, .
\end{equation}
We note that $V^{4,1}\subset\sigma_2^*(V^{4,1})\subset\sigma_3^*(V^{4,1})$ and 
$\tau_2(V^{4,1})\subset\tau_3(V^{4,1})$.
The closed variety $\sigma_2(V^{4,1})$ contains three entanglement types as before, and we have obtained two others, $\sigma_3^*(V^{4,1})\backslash \sigma_2^*(V^{4,1})$ and $\tau_3(V^{4,1})$. The symmetric four-qubit states are classified therefore into three (secant) families ($V^{4,1},\sigma_2,\sigma_3$), a classification that, as presented, can be refined to five families (separable, symmetric superpositions of two, resp. three states, states of type $W_4$ and $\tau_3$, the latter including the state $X_4$). 

\section{Nesting}
Nesting, i.e. that the classification of \( N \)-qubit states carries over to the case of \( N+1 \) qubits, and thus to higher a number of qubits, is a desirable property for entanglement classifications introduced in Ref.~\cite{MEdCSLS15} in order to make it more practical. In fact, from the point of view of applications of quantum information scalability is a crucial property to be assessed, and a resource classification should take this aspect in consideration.

In the classification presented here there is indeed a natural nesting structure, which can be described by a natural identification of 
\(
\sigma^*_k(V^{N,1})\) and \( \sigma^*_{k}(V^{N+1,1})\)
for $k \leq \lfloor \frac{N}{2} \rfloor +1$. This can be constructed by mapping $\sum_{r=1}^k \lambda_r |v_r\rangle^{\otimes N} \leftrightarrow \sum_{r=1}^k \lambda_r |v_r\rangle^{\otimes (N+1)}$: if the set \( \{|v_r\rangle^{\otimes N}\}_{r=1}^k \) is linearly independent, then the set  \( \{|v_r\rangle^{\otimes (N+1)}\}_{r=1}^k \) is also  linearly independent \cite{MEdCSLS15}. Bear in mind, however, that their closures can be different.

This statement (that their closures can be different) is best understood from a nesting property for the tangent varieties,
\begin{equation}
\tau_k(V^{N,1}) \approx \tau_k(V^{N+1,1}),
\end{equation}
with $k\leq \lfloor \frac{N}{2}\rfloor$, as shown for the $W_N$ state. To understand the limitation, first consider \( N=2 \), for which case \( \tau_2(V^{2,1})=\emptyset \), while \( \tau_2(V^{3,1}) \) is non empty. Next, passing to the general case, \(\sigma_{M+1}^*(V^{2M,1})\) is identified with \(  \sigma_{M+1}^*(V^{2M+1,1}) \), while 
\(  \mathrm{dim}( \sigma_{M+1}(V^{2M,1})=2M \) and \(  \mathrm{dim}( \sigma_{M+1}(V^{2M+1,1})=2M+1 \). It follows that \( \tau_{M+1}(V^{2M+1,1}) \) is necessarily larger than \( \tau_{M+1}(V^{2M,1}) \) .

This feature tells us that we can use the classification shown for $N=4$  for partially classifying  $N=5$. To complete this latter case we only need to identify the new SLOCC classes in the tangent variety $\tau_3(V^{5,1})$.
}

\section{Reduced Density Matrices via Sylvester Construction}

We shall now interpret the symmetric tensor border rank introduced in section \ref{b-rank} in terms of reduced density matrices of subsystems. To this end, we introduce the Sylvester construction \cite{Sy52}:
Consider an \( N \)-qubit  vector \( |\psi\rangle\in \mathrm{Sym}\left[\left(\mathbb{C}^2\right)^{\otimes N}\right] \). This vector defines a family of linear transformations, \( \left\{\Psi_j\right\}_{j=1}^{N-1} \), as follows. \( \Psi_j \) is a linear transformation from the dual space $\left\{ \mathrm{Sym}\left[\left( \mathbb{C}^2\right)^{\otimes{j}}\right]\right\}^*$ to the space $\mathrm{Sym}\left[\left( \mathbb{C}^2\right)^{\otimes(N-j)}\right]$, given in physics notation by 
\begin{equation} \label{transfcata}
\Psi_j(\langle\phi_j|)=\langle \phi_j|\psi\rangle .
\end{equation}
In the induced basis of \( \mathrm{Sym}\left[\left(\mathbb{C}^2\right)^{\otimes N}\right] \), the matrix of the $j$-th linear transformation $\Psi_j$ is just the catalecticant matrix 
$C_j(\psi)$ given in \eqref{eq:catalecticant}.
The crucial issue is that the reduced density matrix for \( N-j \) qubits (i.e. after tracing out \( j \) qubits) of state $\psi$ can be written as 

 \begin{equation}
 \rho^{(N-j)}={\Psi}_j\circ{\Psi}_j^{\dag}\,.
 \end{equation}
 This comes about as follows: the reduced density matrix is a linear operation of trace one mapping \( \mathrm{Sym}\left( \mathcal{H}^{\otimes(N-j)}\right) \) to itself, written in physics notation by dualising as \( \langle\varphi_{N-j}| \rho^{(N-j)}|\phi_{N-j}\rangle= \langle\varphi_{N-j}| \mathrm{Tr}_j(|\psi\rangle\langle\psi|)|\phi_{N-j}\rangle\).

It follows that the rank of $\rho^{(N-j)}$ is precisely that of the linear transformation ${\Psi}_j$, and that the classification in \( k \)-secant families is a classification in terms of the ranks of the reduced density matrices. It is important to notice that we are considering the symmetric rank of the reduced density matrices, i.e. their rank as acting on the corresponding symmetric Hilbert spaces; nonetheless it can be computed equally as their rank when acting on the full tensor Hilbert space.
\section{Entanglement Families and Hamiltonians}

A first consequence of this result is that there is a symmetry, in what regards matrix rank, of reduced density matrices of \( N \) spin-$1/2$ particles with respect to the central value  \( j=\lfloor{N/2}\rfloor \).  Secondly, and crucially to provide a physical understanding of the classification in \( k \)-secant families, it provides us with a bound for the interaction length required by a parent Hamiltonian \cite{SWPGC09, MEdCSLS15, HHS10, CSWCPG13} to have this state as a ground state, as follows. If the reduced density matrix for $j$ particles $\rho_j$ possesses a non-trivial kernel $\mathrm{ker} (\rho_j)$, then the projector $P_j$ onto this kernel is non-negative, \( P_j\geq0 \). If we construct the Hamiltonian $H = \sum_{i=1}^N h_i$, with $h_i = \mathbbm{1}^{\otimes i-1} \otimes P_j \otimes \mathbbm{1}^{\otimes N-i-j+1}$, then $H\geq 0$ and $H|\psi_N\rangle =0$, so $|\psi_N\rangle$ is in the ground manifold of $H$. For a $j$-qubit density matrix in the symmetric space, the maximum rank is $j+1$, since that is the dimension of the corresponding symmetric space. In the construction of parent Hamiltonians, we are interested in identifying which is the smallest $j$ such that \( \mathrm{rank}\left( \rho^{(j)} \right)<j+1\), strictly smaller than the maximum possible rank. Clearly, given the symmetry \( \mathrm{rank}\left( \rho^{(j)}\right)= \mathrm{rank}\left( \rho^{(N-j)}\right) \), the interaction length can then always be chosen smaller or equal to \( \lfloor{N/2}\rfloor+1 \) \cite{MEdCSLS15}. 

Nonetheless, we can improve the bounds easily, obtaining much stronger results. By construction, if a given state is in $\sigma_k(V^{N,1})$ but not in $\sigma_{k-1}(V^{N,1})$, then ${\rm rank}\, (\rho^{(j)}) = \min (j,k)$, where the density matrix is understood as a linear operator on the \emph{symmetric} space. As a consequence, it shows a non-trivial kernel for $j =k$, again in the symmetric space, and this bound is tight, which means that this is the minimal possible interaction length for a translation-invariant frustration-free Hamiltonian to have this state as a ground symmetric state. This bound improves drastically the one obtained in Ref. \cite{MEdCSLS15}. This may be illustrated with the $W_N$ state, which is always in $\sigma_2(V^{N,1})$, so the predicted interaction length is $j = 2$, compared to $j \leq \lfloor \frac{N}{2}\rfloor +1$, and indeed, in Ref.~\cite{MEdCSLS15} we provided a parent Hamiltonian with interaction length $2$ for all $N$.

\section{Discussion and Conclusions}

The relationship between entanglement theory for pure states and algebraic geometry is extremely deep. For instance, although we have limited ourselves to the symmetric case and Veronese secant varieties, there is a natural extension for the general case of qudits in terms of Segr\'e embeddings and the corresponding secant varieties. The Segr\'e embedding is given by \( \left([\psi],[\phi]\right)\to[\psi\otimes\phi] \) in the bipartite case (for a geometric application in entanglement see for instance \cite{Brody200119}), and identifies the class of separable states in any bipartition as a determinantal variety. Furthermore, the problem of classification of entanglement in the multipartite case corresponds to the old problem of the classification of orbits of the projective group \( \mathbb{P}\left( \mathrm{GL}\left(\mathcal{H}_1\right)\times\mathrm{GL}\left(\mathcal{H}_2\right)\times \cdots \times\mathrm{GL}\left(\mathcal{H}_n\right)\right) \) in the projective space \( \mathbb{P}\left( \mathcal{H}_1\otimes \mathcal{H}_2\otimes\cdots\otimes \mathcal{H}_n\right) \). In this vein, the six SLOCC classes  of three qubits presented in \cite{DVC00} give the full answer to the problem in that case. Several authors have already used the approach of algebraic geometry to advance towards a full SLOCC classification, 
see in particular 
\cite{PhysRevA.67.012108} and 
\cite{jmp10.1063/1.4753989}, 
and it remains a fruitful avenue to explore further.

We have proposed the study of multipartite entanglement classification in the language of algebraic geometry. In this line of thought we have used our results to illuminate the connection between the \( \mathrm{GHZ}_N \) and \( W_N \) classes in more physical terms: the secant level to which a state first belongs is equal to the minimum interaction length of its parent Hamiltonians. Finally, we believe that algebraic geometry will provide novel tools and a fresh view to the intricate problems of entanglement in quantum information~\cite{HJN16}.

The authors acknowledge support from Spanish MINECO FIS2015-69983-P, UPV/EHU UFI 11/55, and TRR80 of the Deutsche Forschungsgemeinschaft.

\end{document}